\documentclass[11pt]{article}

\usepackage[T1]{fontenc}
\usepackage[utf8]{inputenc}

\usepackage{url}
\usepackage{epsfig}
\usepackage{adjustbox}
\usepackage{indentfirst}
\usepackage{float}
\usepackage{xcolor}
\usepackage{makecell}
\usepackage{amsmath}

\title{Cost models for geo-distributed massively parallel streaming analytics}
\author{
        Anna-Valentini Michailidou and Anastasios Gounaris \\
                Aristotle University of Thessaloniki, Greece\\
                \{annavalen,gounaria\}@csd.auth.gr
            \and
        Konstantinos Tsichlas\\
        University of Patras, Greece\\
                    ktsichlas@ceid.upatras.gr
}
\date{}

\begin{document}
\maketitle

\pagenumbering{arabic}
\setcounter{footnote}{1}

\begin{abstract}
This report is part of the DataflowOpt project on optimization of modern dataflows and aims to introduce a data quality-aware cost model that covers the following aspects in combination: (1) heterogeneity in compute nodes, (2) geo-distribution, (3) massive parallelism, (4) complex DAGs and (5) streaming applications. Such a cost model can be then leveraged to devise cost-based optimization solutions that deal with task placement and operator configuration.

\end{abstract}

\section{Introduction}
A fundamental research activity with a view to devising new methodologies regarding advanced dataflow (i.e., data-intensive workflow) optimization solutions is to develop and adopt appropriate cost models. Such cost models need to be capable of accurately reflecting the response time in parallel environments, taking into account issues such as partial overlapping, CPU core sharing and resource contention that are all common in modern dataflow executions. Furthermore, the cost models need to be extended to cover additional dataflow objectives, such as data quality. Finally, these cost models need to cover arbitrarily complex DAG (directed acyclic graph) execution plans. 
This report, apart from examining cost models that are directly amenable to optimizations, further discusses tractability issues regarding the optimization problems that typically accompany cost models. 

More specifically, in Section \ref{sec:CostModels}, we present alternative approaches to develop cost models with a view to covering the landscape that is characterized by the  combination of (1) heterogeneity in compute nodes, (2) geo-distribution, (3) massive parallelism, (4) complex DAGs and (5) streaming applications.   We then highlight the need to develop our own cost model, which covers the afore-mentioned aspects, i.e., heterogeneity, massive parallelism and geo-distribution, while being suitable for data-intensive streaming applications. This cost model, is presented in detail in Section \ref{sec:our-cost-model} and the relevant code is in \url{https://github.com/annavalentina/Equality}.\footnote{This code repository contains broader work.}
Section \ref{sec:summary} concludes the report.

\section{Cost Models for Dataflow Optimization}
\label{sec:CostModels}

In this section, we discuss the main representatives regarding the cost models for data intensive analytics, along with the associated optimization problems, whenever such problems accompany a cost model presentation, and tractability analysis. The aspects that are of immediate interest include:  massive parallelism, heterogeneity in terms of resources and (geo-)distribution, while we assume that the complete job is modelled by a (potentially complex) DAG. Please note that we use the terms DAG, topology, workflow (or dataflow) interchangeably in the text.

\subsection{A Cost Model for a Multi-core Shared Memory Setting Tailored for Streaming Data Analytics}

Zhang et al. \cite{ZhangHZH19} propose a cost model that is used to maximize the processing throughput of data stream processing systems on shared-memory multi-core architectures. The throughput $R$ is equal to the sum of the output rates of all the sink nodes of the topology, that is $R=\sum_{sink}{r_{o}}$. The output rate $r_{o}$ for a time interval $t$ is estimated using the number of tuples that are processed, $num$, and the time needed to process them $t_{p}$. Hence $r_{o}=num/t_{p}$. $num$ is equal to the number of tuples generated from the previous nodes in the topology, $num=\sum_{producers}{r_{i}*t}$ where $r_{i}$ is the input rate. The $t_{p}$ variable is equal to $t_{p}=\sum_{producers}{r_{i}*t*T(p)}$ where $T(p)$ is the average time spent on handling each tuple and consists of the fetching ($T^{f}$) and execution ($T^{e}$) times. The fetching time $T^{f}$ is 0 in case the producer keeps its data locally; otherwise it is equal to $[N/S]*L_{(i,j)}$ where $N$ is the average size of a tuple, $S$ the cache line size and $L_{i,j}$ the worst case memory access latency between two CPU sockets $i$ and $j$ in a generic NUMA computing architecture. The variables $L_{i,j}$ and $S$ are given as statistical input metadata, whereas $N$ and $T^{f}$ are found using profiling techniques. Finally, the input rate of the source operators is also given by the problem setting (denoted as $I$). The algorithms proposed in \cite{ZhangHZH19} aim to maximize the throughput $R$ by deciding a placement of nodes to CPU sockets and a replication level for each operator. The proposed algorithms focus on the degree of parallelism of the operators taking into account the heterogeneity of the resources, without considering geo-distribution.  Apart from this difference from  our main focus, the heterogeneity in \cite{ZhangHZH19} is limited and refers to the locality or not of data in a NUMA architecture, which impacts on the fetching time as described above.

In summary, this cost model emphasizes on the different fetching times based on the data memory address and the placement of the operator.

\subsubsection{Optimization Problem Details and Tractability}

Zhang et al. \cite{ZhangHZH19} aim to maximize the application processing throughput, given an input stream rate by placing each operator on specific CPU sockets $S_i$, where $i=1\dots n$, and deciding an optimal replication level in a multicore shared-memory architecture. The formulation of the problem they try to solve is the following:

\begin{equation*}
maximize\; \sum_{sink}\overline{r_o} (i.e.,~ expected ~r_o)
\end{equation*}
\begin{equation*}
s.t., \forall i,j \\\ \epsilon \\\ 1,...,n,
\end{equation*}
\begin{equation}\label{zhang1}
\sum_{operators\\\ at\\\ S_i}\overline{r_o} * T \leq C,
\end{equation}
\begin{equation}\label{zhang2}
\sum_{operators\\\ at\\\ S_i}\overline{r_o} * M \leq B,
\end{equation}
\begin{equation}\label{zhang3}
\sum_{operators\\\ at\\\ S_j}\sum_{producers\\\ at\\\ S_i}\overline{r_o(s)} * N \leq Q_{i,j}.
\end{equation}

Equation (\ref{zhang1}) ensures that the aggregated CPU demand on each socket will not exceed its CPU capacity $C$, where $T$ is the average time spent on handling each tuple. Similarly, equation (\ref{zhang2}) ensures that the aggregated amount of bandwidth requested on a socket (calculated using the average memory bandwidth consumption per tuple, $M$) will not exceed its maximum attainable local DRAM bandwidth $B$. Finally, equation (\ref{zhang3}) enforces that the data transfer from one socket to another must not exceed the maximum remote channel bandwidth. It is also constrained that each operator is placed exactly once. 

Each operator of the total $|o|$ operators can be replicated at most $k$ times, thus the total possible configurations are $k^{|o|}$. For each of these configurations, there exist $m^n$ different placements, where $m$ is the number of sockets and $n$ the number of replicas. Due to the exponential search space and the corresponding ILP that needs to be solved, a brute-force approach is prohibitive. The authors propose a branch and bound algorithm that iteratively places operators under a given replication level and then tries to increase the degree of replication for the bottleneck operator. In total, there exist $(n*m)^n$ candidate solutions ($n*m$ for $n$ iterations) and three heuristics are used in order to speed up the traversal of the solution space.

\subsection{Cost Models for Data Flows Tailored to Parallel Executions}
Kougka et al. \cite{KougkaG19} present three cost models for data flow response time in a massively parallel environment taking into account execution overlaps. Their approach can be extended to parallel distributed settings but heterogeneity of the resources is not considered. 

The first two models refer to pipelined segments where the tasks in the data flow DAG form a chain and, also, no movement of data over the network occurs, since the execution takes place on a single machine (thus making communication cost time negligible).   The idea behind the first cost model is that each task can be executed on a separate core resulting to overlapping execution times. In such a case, the slowest execution time defines the total response time. Also the  authors in  \cite{KougkaG19} account for overheads coming from factors such as memory locks, contention and context switching between threads, where all these factors are captured  by a single variable $\alpha$. The model proposed is the following: $Response Time(RT)=\alpha*max\{c_{1},...,c_{n}\}$, where $c_{i}$ is the execution time of task $i$ in the DAG. 

In the second model, the case where a core needs to take on more than one task is examined. The new model is the following: $ResponseTime(RT)=\alpha*max\{max\{c_{1},...,c_{n}\},\sum\{c_{1},...,c_{n}\}/m\}$ where $m$ is the total number of cores. The outcome of this model is the same as the first one in case there exists a single task with execution time higher than the sum of the execution times of all the tasks. 

Finally, a generalized model is presented that takes into account multiple pipeline segments and multiple machines. The model presented is the following: $Response Time(RT)=\sum z_{i}*w^{c}*c_{i} + \sum z_{ij}*w^{cc}*cc_{i \rightarrow j}$. The first sum refers to execution time whereas the second one to communication cost time. The $z$ variables are binary ones that are equal to 1 when the task and the communication costs contribute to the final $RT$, capturing the overlapping of tasks. The $w$ variables generalize the $\alpha$ variable used in the previous models and capture the possible overheads of the execution. Finally, $cc_{i \rightarrow j}$ represents the communication cost of the edges between the tasks. 

In summary, the cost models  in  \cite{KougkaG19}  emphasize on capturing the effect of concurrent task executions on the response time in a parallel homogeneous setting.

In a similar setting, Ali et al. \cite{AliW19} propose a cost model for optimizing User Defined Functions (UDFs) in data-intensive workflows, such as ETL (Extract-Transform-Load) workflows. They try to optimize the execution of the UDFs in ETL workflows based on the execution time and monetary costs of task invocations. The optimization is divided into three stages. In the first one, the cost model determines if a UDF can be parallelized and in the second stage, it determines the degree at which  it should be parallelized. Finally, in the third stage, the cost model generates an efficient machine configuration. The cost model is used to simulate the UDF execution  and derive estimates regarding execution time as well as other metrics like the dataset size, the execution speed, CPU, I/O, memory characteristics etc. 

In summary, the cost model in \cite{AliW19} focuses on the impact of increased degree of parallelism regarding certain expensive operators in the DAG.

\subsubsection{Optimization Problem Details and Tractability}

The cost model in  \cite{KougkaG19} mentioned above is leveraged to find an optimal (partial) ordering of the operators (nodes) in  a DAG, in the case that multiple orderings can produce the same result. The response time optimization in such a setting is not simpler than the general task ordering optimization problem aiming at the minimization of the sum of the costs, as reported in \cite{KougkaGS18}. This problem, as proven in \cite{BMS05}, is intractable and more importantly \textit{``it is unlikely that any polynomial time algorithm can approximate the optimal plan to within a factor of $O(n^\theta)$''}, where $\theta $ is some positive constant and $n$ is the number of the tasks in the data flow.

Ali et al. \cite{AliW19}, as already explained, optimize the execution of UDFs in ETL workflows using also the sum cost metric. The workflows are divided into stages and each stage can be processed using multiple code/implementation variants. The objective is to find the optimal variant for each stage. In total, there exist $n^m$ possible variants ($m$ stages, $n$ code variants for each stage). This problem is NP-hard since the Multiple Choice Knapsack Problem (MCKP) is a special case of it. In the MCKP, given $m$ stages $N_1,...,N_m$ of the workflow to pack in a knapsack of capacity $B$ (the budget constraint), where each code variant $j\; \epsilon \; N_i,i=1,2,...,m$, has execution time $T_{i,j}$ and cost $C_{i,j}$, the problem is to choose exactly one code variant from each stage such that the sum of execution times is minimized while the sum of costs does not exceed the budget $B$. The proposed formulation is the following (the decision variable is $x_{ij}$): 

\setcounter{equation}{0}
\begin{equation*}
minimize\;(Z) \qquad \sum_{i=1}^m \sum_{j \epsilon N_i} T_{i,j}*x_{i,j}
\end{equation*}
Subject to:

\begin{equation}\label{ali1}
\sum_{i=1}^m \sum_{j \epsilon N_i} C_{i,j}*x_{i,j} \leq B,
\end{equation}
\begin{equation}\label{ali2}
 \sum_{j \epsilon N_i}x_{i,j}=1, i=1,...,m,
\end{equation}
\begin{equation}\label{ali3}
x_{i,j} \; \epsilon \; \{0,1\}, j \; \epsilon \; N_i, i=1,...,m.
\end{equation}

\subsection{Cost Models Tailored for Streaming Data Analytics in an Edge/\\Fog/Cloud Computing Setting}

Hiessl et al. \cite{HiesslKHSN19} aim to minimize response time, migration and enactment cost as well as to maximize availability in data stream processing systems by placing (and replacing) operators of stream topologies. However, the operator execution is not parallelized, i.e., each DAG node is placed on exactly one compute node. Their proposal targets fog environments, where heterogeneity and geo-distribution of the resources are inherent characteristics and they extend the system model and formulation presented in \cite{CardelliniGPN16}. The adopted cost model represents both the resources and operators as graphs. In the former case, the links represent the network links where,  in the latter case, the graphs refer to the flow of data streams between the operators. Each compute node has its CPU frequency, number of CPU cores, memory and storage capacity, whereas each operator has its own CPU frequency, number of CPU cores, memory and storage requirements. Also, for each operator, there exists a subset of the resources that are available to allocate its execution. The operators can be enacted and migrate from one compute node to another.  The response time of a topology is equal to the maximum delay of all paths in the stream data flow, which depends on the processing time of tuples and on the delay of sending the tuples. The modeling includes the $C_{op}(x)$ variable, which refers to the total enactment cost per second,  which is the cost of running the topologies on the available resources. Another variable is $C_{mig}(x)$, which is the cost that comes from all the migrations of operators from one compute node to another and depends on the operator image size as well as the data rate of pulling this operator to a compute node.

In general, this model focuses on capturing the environment heterogeneity and then, in the subsequent optimization problem, the emphasis is placed on expressing several constraints.

Renart et al. \cite{RenartVBALP19} also place operators across edge and cloud devices in order to optimize metrics like end-to-end latency, WAN traffic and messaging cost. The setting they work on is geo-distributed with heterogeneous resources but the operators are not parallelized, as in the cost model above. Each resource is characterized by its CPU and memory capabilities ($cpu,mem$). Each network link $l_{k \leftrightarrow l}$ has a certain bandwidth capability $bdw_{k \leftrightarrow l}$ with other resources. Each operator has its own CPU and memory requirements ($cpu, mem$), selectivity, process rate $\mu_{<i,k>}$, input and output size $\varsigma^{in}_{i}, \varsigma^{out}_{i}$ and output production rate $\lambda^{out}_{i}$, which is based on its input rate $\lambda^{in}_{i}$. The resource and operator modeling is based on an M/M/1 queue model, which allows for calculating the computation and communication times.
$stime_{<i,k>}=\frac{1}{\mu_{<i,k>}-\lambda^{in}_{i}}$ gives the computation time of an operator $i$ placed on a resource $k$ and 
 $ctime_{<i,k><j,l>}=\frac{1}{(\frac{bdw_{k \leftrightarrow l}}{\varsigma^{out}_{i}})-\lambda^{in}_{j}}+l_{k \leftrightarrow l}$ denotes the communication time for an operator $i$ placed on resource $k$ to send data to operator $j$ placed on resource $l$.
The end-to-end latency $L_{p_{i}}$ of a path $p_i$ is the sum of the computation times of all the operators in this path plus the communication times between them. The way the latency of each path is calculated in \cite{RenartVBALP19} is similar to that of \cite{HiesslKHSN19} but the latter focuses on optimizing the maximum latency of all paths, whereas this work focuses on the sum of the latencies. The WAN traffic of a path $W_{p_{i}}$ indicates the size of messages that cross it. The messaging cost of a path $C_{p_{i}}$ shows the number of messages between edge and cloud. Finally the total aggregate cost model they use is the following:
 $AggregateCost_{p_{i}}=w_{l}*\frac{L_{p_{i}}}{Par_{lat}}+w_{w}*\frac{W_{p_{i}}}{Par_{wan}}+w_{c}*\frac{C_{p_{i}}}{Par_{cost}}$, where $w_{l}+w_{w}+w_{c}=1$ are weights and $Par_{lat},Par_{wan},Par_{cost}$ indicate the current latency, WAN traffic and messaging cost of the running application.
 
 This cost model is characterized by the usage of queue theory and the fact that it aims to capture the environment heterogeneity, which is a main goal for cost models in geo-distributed settings. As previously, the final cost is a weighted sum of multiple individual costs, which are computed by simple summation expressions, given that these cost models are typically leveraged by multi-objective placement optimization solutions.
 
\subsubsection{Optimization Problem Details}

Hiessl et al. \cite{HiesslKHSN19} place stream operators in fog computing environments. Both the computing resources and the application are presented using topology graphs. The aim is to map the operators and edges from the application graph to the resources and network links of the resource graph, respectively, in order to minimize the following objective function: $F'_{cost}=F'(x,y,r)+w_{c_{op}}\frac{C_{opmax}-C_{op}(x)}{C_{opmax}-C_{opmin}}+w_{c_{mig}}\frac{C_{migmax}-C_{mig}(x)}{C_{migmax}-C_{migmin}}$, where $F'(x,y,r)=w_{r}\frac{R_{max}-r}{R_{max}-R_{min}}+w_{a}\frac{logA(x,y)-logA_{min}}{logA_{max}-logA_{min}}$. The $w$ variables are weights used in the simple additive weighting method that applies to multi-objective problems. The $F'(x,y,r)$ variable focuses on response time and availability ($R$ and $A$ respectively).

The formulation of the problem is the following:

\setcounter{equation}{0}
\begin{equation*}
minimize\; F'_{cost}
\end{equation*}
Subject to:

\begin{equation}\label{hiessl1}
C_{op}(x)\leq B_{op}
\end{equation}
\begin{equation}\label{hiessl2}
C_{mig}(x)\leq B_{mig}
\end{equation}
\begin{equation}\label{hiessl3}
\sum_{u\epsilon V_{res}} T_{(actual,i)} \frac{S_i}{S_u} x_{i,u} \leq T_{(max,i)} \qquad \forall i \; \epsilon \; operators
\end{equation}
\begin{equation}\label{hiessl4}
\sum_{i\epsilon V_{dsp}} P_{(CPU,i)} P_{(Cores,i)} x_{i,u} \leq  P_{(CPU,u)} P_{(Cores,u)} \qquad \forall u \; \epsilon \; resources
\end{equation}
\begin{equation}\label{hiessl5}
\sum_{i\epsilon V_{dsp}} P_{(Mem,i)} x_{i,u} \leq  P_{(Mem,u)} \qquad \forall u \; \epsilon \; resources
\end{equation}
\begin{equation}\label{hiessl6}
\sum_{i\epsilon V_{dsp}} P_{(HD,i)} x_{i,u} \leq  P_{(HD,u)} \qquad \forall u \; \epsilon \; resources
\end{equation}
\begin{equation}\label{hiessl7}
r \geq \sum_{k=1}^{n_p}\sum_{u\epsilon V^{i_k}} \frac{ET_i}{S_u} x_{i_k,u}+ \sum_{k=1}^{n_p-1}\sum_{(u,v)\epsilon \;available\_ resources} d_{(u,v)} y_{(i_k,i_{k+1}),(u,v)} \qquad \forall  p \; \epsilon \; paths
\end{equation}

Equations (\ref{hiessl1}) and (\ref{hiessl2}) limit  the enactment and migration cost using budget constraints $B_{op}$ and $B_{mig}$, respectively. Equation (\ref{hiessl3}) ensures that the processing duration of an operator assigned to a resource should not exceed the maximum processing time limit; in the equation, $S_i$ denotes the speed of processing of operator $i$ in previous optimization circles and $S_u$ the speed of processing of resource $u$.
Equations (\ref{hiessl4}), (\ref{hiessl5}) and (\ref{hiessl6}) ensure that the CPU, Memory and storage requirements of the operators are fulfilled while not exceeding the capacities of the resources. Finally, equation (\ref{hiessl7}) ensures that the total response time is greater than or equal to the response times of all the topology paths; $d$ denotes the network delay and $ET$ the execution time per tuple,
To avoid quadratic or higher order equations, the authors skip the multiplication of the decision variables $x_{i,u}$ and $y_{(i,j),(u,v)}$ which refer to the placement of operators and links, respectively.

Renart et al \cite{RenartVBALP19} place operators of IoT dataflows on either edge or cloud resources in order to minimize end-to-end latency, data transfer and messaging cost between the edge and the cloud. The aggregate cost formula for a single path has been presented above. The formulation of the problem they try to solve that is given below refers to the aggregation costs across all paths:

\setcounter{equation}{0}
\begin{equation*}
minimize\; \sum_{p_i\epsilon P}AggregateCost_{p_i}
\end{equation*}
Subject to:

\begin{equation}\label{renart1}
\lambda^{in}_{o} < \mu_{<o,r>} \qquad \forall o \; \epsilon \; operators,  \forall r \; \epsilon\; resources|mo_{<o,r>}=1
\end{equation}
\begin{equation}\label{renart2}
\lambda^{in}_{o} < (\frac{bdw_{k \leftrightarrow n}}{\varsigma^{out}_{o-1}}) \qquad \forall o \; \epsilon\; operators,  
\forall k \leftrightarrow n \; \epsilon \; links|mo_{<o,k>}=1
\end{equation}
\begin{equation}\label{renart3}
\sum_{o \epsilon O}mo_{<o,r>} * \lambda^{in}_o \leq cpu_r \qquad \forall r\;\epsilon\; resources
\end{equation}
\begin{equation}\label{renart4}
\sum_{o \epsilon O}mo_{<o,r>} *mem_{<o,r>} \leq mem_r \qquad \forall r\; \epsilon \;resources
\end{equation}
\begin{equation}\label{renart5}
\sum_{\substack{s_i \rightarrow \epsilon S \\   k \leftrightarrow l \epsilon L}}ms_{<i \rightarrow j, k \leftrightarrow l>} *\varsigma^{out}_i \leq bdw_{k \leftrightarrow l}
 \qquad \forall k \leftrightarrow l \; \epsilon \; links
\end{equation}
\begin{equation}\label{renart6}
\sum_{r \epsilon R}mo_{<o,r>}=1 \qquad \forall o\; \epsilon \; operators
\end{equation}
\begin{equation}\label{renart7}
\sum_{k \leftrightarrow l \epsilon L}ms_{<i \rightarrow j ,k \leftrightarrow l>}=1 \qquad \forall s_{i \rightarrow j}\; \epsilon \; streams
\end{equation}

The variable $mo_{<o,r>}$ is set to 1 if operator $o$ is placed on resource $r$ and, similarly, $ms_{<i \rightarrow j, k \leftrightarrow l>}=1$ if the stream between operators $i,j$ is assigned on the link between resources $k,l$. Equations (\ref{renart1}) and (\ref{renart2}) ensure that the input rate does not exceed the process rate of the resource and that the link is not saturated, respectively. Equations (\ref{renart3}) and (\ref{renart4}) ensure that the CPU and memory requirements of each operator are fulfilled and the resource capacities are not exceeded. Equation (\ref{renart5}) ensures that the bandwidth limits of the links are not exceeded. Finally, equations (\ref{renart6}) and (\ref{renart7}) ensure that an operator is placed on only one resource and that each data stream is placed on only one link, respectively.

\subsubsection{Tractability}

In both \cite{HiesslKHSN19} and \cite{RenartVBALP19}, the problem is more complex than the problem of choosing among multiple execution engines, when the placement decision of a node affects the other placements. This problem is analyzed in \cite{KougkaGT15}, and is proven that it is NP-hard, and moreover, \emph{cannot be approximated by a polynomial-time algorithm with approximation error bound less than 8/7}.
Also, evidence about the NP-hardness is provided in \cite{CardelliniGPN16}. As shown from the problem formulations, such task placement problems are specific forms of constrained optimizations (mixed) integer linear programming problems.

\subsection{A Cost Model Tailored to Stage-by-stage Analytics in a Cloud Computing Setting } 
 
Gounaris et al. \cite{GounarisKNM14} propose a bi-objective cost model in order to optimize queries in multi-cloud environments. The queries are in a form of DAGs and are divided into strides where each stride represents a step in the execution and consists either of one operator or a group of operators running in parallel. However each operator cannot be parallelized. The goal is to assign strides to cloud-based VMs. Each cloud provider offers VMs to users and charges  based on one of the following policies : on-demand, where the payment is computed based on the exact usage time, reserved, where the user pays upfront for a specific time and then pays on-demand, spot instance, which is a bidding process or committed, where the customer rents the VM for a predefined time. The model they present estimates the execution time and monetary cost and models the charging policies and fees. The basic formula for modeling the execution time is the following:
$TotalTime=\sum_{s=1}^n max(S_{s,1 \rightarrow i}^{VM_{k \rightarrow k'}},...,S_{s,m^s \rightarrow j}^{VM_{k \rightarrow k'}})$,
where $n$ is the number of strides, $m^s$ the number of operators in the $s^{th}$ stride and $S_{s,i \rightarrow j}^{VM_{k \rightarrow k'}}=O_{s,i}^{VM_{k}}+T_{s,i \rightarrow j}^{VM_{k \rightarrow k'}}$. $O_{s,i}^{VM_{k}}$ is the time it takes to execute the $i^{th}$ operator of the $s^{th}$ stride on $VM_k$ and depends on the hardware characteristics of the heterogeneous VMs. $T_{s,i \rightarrow j}^{VM_{k \rightarrow k'}}$ is the time to transfer data produced from operator $i$ which runs on $VM_k$ to operator $j$ on $VM_{k'}$. In case there exist network bottlenecks, the total time is not the max of the sum of execution and transmission time of each operator but rather the sum of them as follows:
$TotalTime=\sum_{s=1}^n \sum_{i=1}^{m^s} S_{s,i \rightarrow j}^{VM_{k \rightarrow k'}}$. In case pipelining is supported, an operator can start transmitting data before its execution is complete. Then the next formulation holds $S_{s,i \rightarrow j}^{VM_{k \rightarrow k'}}=max(O_{s,i}^{VM_{k}},T_{s,i \rightarrow j}^{VM_{k \rightarrow k'}}$). 
The monetary cost of a query is the following:
$\sum_{s=1}^n \sum_{i=1}^{m^s} Price(S_{s,i \rightarrow j}^{VM_{k \rightarrow k'}},F_{pr}(P_t^\alpha,VM_k),F_{pr}(P_{t'}^\alpha,VM_{k'}))$, where $F_{pr}$ represents the price offer from cloud provider $pr$ for $Vm_k$ according to the pricing policy $P_{t}^\alpha$. Then Price finds the fee for using $VM_k$ for $S$ time and transferring data to $VM_{k'}$.

\subsubsection{Optimization Problem Details and Tractability}
 
 This cost model accompanies bi-objective problems initially encountered in federated databases, where an execution plan is processed stage-by-stage across multiple sites, which are shown to be NP-hard \cite{PY01,tsamoura2013multi}.

\subsection{A Cost Model for (Incremental) MapReduce Jobs}

Li et al. \cite{LiDS15} propose a system built on top of incremental Hadoop that allocates resources and uses runtime scheduling techniques in order to meet user latency requirements while maximizing throughput. The queries are represented using DAGs where each node is either a data distributor or a Map Reduce pair. They model the mean and the variance of latency both per-tuple and per-window. Regarding the per-tuple latency, G/G/1 queues are used to represent streams. The total latency and variance are a sum of  latencies and variances due to 12 causes including batching, queuing, heartbeats awaiting, CPU, network and Disk I/O. The mean CPU latency per tuple is formulated as follows:
\\$E(\ddot{L}_{cpu})= \frac{u}{2*min(1-p,1/n)*C}$, where $u$ is the total resource required by a batch and depends on the average number of tuples and the average resource required per tuple. Also, $p$ is the fraction of the resource required by the other threads on the same node since resources receive multiple map-reduce tasks, $n$ is the number of cores of a CPU and $C$ is the number of processing cycles of a CPU per unit time. Network and Disk I/O are modeled in a similar way. The queuing delay can be calculated based on the theory of the G/G/1 queues. The per-window latency is modeled in cases where the results of a window are computed only when a reducer has received all the tuples and is equal to $E(\tilde{L})=E(U)+E(F)$ where $U$ is the maximum of all the tuples' latencies in this window and F represents the execution of a particular partitioned window.

In summary, this cost model focuses on splitting the latency factors of a tuple processed by a MapReduce round into 12 parts, and takes into account resource sharing.

\begin{table}[tb!]
\centering
\footnotesize
\begin{tabular}{|>{\centering\arraybackslash}p{0.1\textwidth}|>{\centering\arraybackslash}p{0.18\textwidth}|>{\centering\arraybackslash}p{0.12\textwidth}|>{\centering\arraybackslash}p{0.23\textwidth}|>{\centering\arraybackslash}p{0.23\textwidth}|}
\hline
Work         & Setting                               & Objective             & Resources' metadata required &                                                                                    Operators' metadata required \\ \hline \hline
Zhang et al. \cite{ZhangHZH19} & Multicore shared-memory architectures: decide placement of operators to sockets and degree of parallelism & Processing Throughput & CPU cycles per socket, DRAM bandwidth,  bandwidth between sockets, memory access latency, cache line size& Memory bandwidth consumption, execution time,  average size of tuples, input stream rate \\ \hline

Kougka et al. \cite{KougkaG19}  & Distributed multicore machines: decide operator ordering & Response Time         & Communication cost, number of cores                                                                                                                                        & Execution cost, selectivity                                                                                                              \\ \hline

Hiessl et al. \cite{HiesslKHSN19} & Edge/Fog environment: decide placement of operators on compute nodes & Response Time, availability, migration and enactment cost& Availability, delay of network links, speedup, CPU frequency, number of CPU cores, memory and storage capacity per compute node & Speedup from previous run, image size,  requirements for CPU frequency, number of CPU cores,  memory and storage per operator \\ \hline

Renart et al. \cite{RenartVBALP19}& Edge/cloud environment: decide placement of operators on compute nodes& End-to-end latency, WAN traffic, messaging cost& CPU and memory capacities, bandwidth costs & CPU and memory requirements, selectivity, input/output event rate, input/output event size \\ \hline

Gounaris et al. \cite{GounarisKNM14} & Multi-cloud environments: decide placement of operators on compute nodes & Execution time, monetary cost& Charging policy, CPU, I/O access speed & Size of data \\ \hline
Ali et al. \cite{AliW19} & Multicore machines: decide degree of parallelism & Execution time, monetary costs& Configuration, row and byte process rate& Size of data, memory, I/O and CPU usage \\ \hline
Li et al. \cite{LiDS15} &Multicore machines running MR: decide degree of parallelism (amount of resources) and scheduling &  Latency, throughput  &  CPU, disk I/O and network characteristics  &Estimated data rate, resource required and input-to-output ration\\ \hline
\end{tabular}
\caption{Summary of techniques examined.}
\label{tab:summary}
\end{table}

\subsubsection{Optimization Problem Details and Tractability}
The work in Li et al. \cite{LiDS15} decides the amount of resources per Map/Reduce solving a non-linear constrained optimization problem (using MinConNLP). In the generic case, these problems are intractable. The main focus is on scheduling the tuples through re-ordering their execution on-the-fly.

\subsection{Summary of Techniques}

We summarize the techniques examined thus far in Table \ref{tab:summary}. We also provide an overview of their main limitations, which call for the development of a new cost model, tailored to massively parallel complex streaming analytics jobs over heterogeneous geo-distributed nodes; this cost model is presented in Section \ref{sec:our-cost-model}. The cost models in \cite{ZhangHZH19,KougkaG19,AliW19} do not cover  heterogeneity in compute node characteristics and/or locations. The models in \cite{HiesslKHSN19,RenartVBALP19,GounarisKNM14} do not account for massive parallelism. Therefore, they are inadequate and unsuitable to be adopted for our case, which focuses on complex DAGs that are executed over geo-distributed heterogeneous resources while benefiting from partitioned parallelism. 

\subsection{Additional Issues}

We have already identified that the existing cost models fall short in supporting the use cases targeted hereby. For completeness, in this section, we examine more cost modelling approaches and we also consider scheduling problems. 

\subsubsection{Other Modeling Approaches}

Karnagel et al. \cite{KarnagelHL17} estimate the performance of tasks in order to achieve an optimal adaptive placement of OLAP queries on heterogeneous hardware. To estimate the query runtime, they use an automated black-box online-learning approach. They use monitoring for execution times as well as input data cardinalities. To estimate the transfer costs, they rely on statistics collected in the calibration phase rather than performing monitoring. In general, learning task costs using (machine)-learning approaches is quite common and is orthogonal to how individual costs are combined to estimate the cost of the complete DAG.  

Learning methods can also apply to the whole DAG. More specifically,
Witt et al. \cite{WittBGL18} survey predictive performance modeling (PPM) techniques that estimate metrics like execution time, required memory and predict future wait times of tasks and jobs in data-parallel and distributing systems. They compare the works by the performance factors they consider; that is workload (resource usage patterns), heterogeneity (of resources), scale (amount of resources and problem size) and contention (caused by resource sharing). They study both task performance models and job performance ones. The representative works we present from this survey here consider the performance factors of heterogeneity and scalability. In the first category (namely, task performance models), a single task runs on a single node. Marin et al. \cite{MarinG04} predict the performance of tasks, the cost of the computation and the execution duration using memory access patterns that are monitored using histograms, the cache size and input size. Matsunaga et al. \cite{MatsunagaF10} predict execution duration, output file size and resident set size of a task using regression trees and characteristics like CPU speed, cache size, available memory and disk capabilities, number of threads as well as location and size of input. Iverson et al. \cite{IversonOP99} predict the execution duration of a task using the input size and hardware characteristics and a local learning approach. The second category refers to approaches that deal with multiple nodes that comprise a job. Delimitrou et al. \cite{DelimitriouK14} predict the performance by observing other applications with similar resource usage patterns and make use of Singular Value Decomposition to classify incoming jobs. Shanjay et al. \cite{SanjayV08} use regression to model the computation duration and the non-overlapped communication duration, whereas the total execution duration is equal to their sum. Cunha et al. \cite{CunhaRTN17} aim to predict performance in order to choose between cloud and a local cluster. To achieve this, they propose a linear model and an empirical one that favors local execution when the prediction is unclear. Last, Alipourfard et al. \cite{AlipourfardLCVYZ17} propose an off-line search method that uses a Gaussian model to select cloud configurations like number of virtual machines, number of cores, CPU speed, RAM per core etc in order to optimize the performance of a job. 
 
Finally, the work of Agrawal et al. \cite{AgrawalBDR12} selects the task ordering and the assignment of tasks to compute nodes, while it targets both the bottleneck cost metric (called period) and the critical path cost (which is equivalent to the latency); this work considers precedence constraints between the tasks in the DAG but the exact ordering is flexible. Each task has a cost and a selectivity (selectivities are independent). Also, there is a set of servers acting as the compute nodes. 
The focus is on both homogeneous and heterogeneous platforms (heterogeneity refers to network speeds) but not on geo-distributed settings. The  communication speed is considered in some flavors. If there are no communication costs, for the cases of homogeneous machines,  Agrawal et al. present algorithms that perform optimal ordering in polynomial time. They also prove that if the machines are  heterogeneous, the problems are NP-hard even when no precedence constraints exist. On the other hand, if there are communication costs, they consider the models of overlapping (communication fully overlaps with computation), and sequential (no overlap) processing. The problem is always NP-complete, apart from the case of period minimization and overlap. In general, in this work the cost models are simple and the value lies in the algorithmic analysis and solutions, which however refer to a more restricted setting than the one assumed in this report.

\subsubsection{Cost Models in Scheduling Works}

We organize the discussion in this section around five recent surveys in the area of scheduling.

First, Hong et al. \cite{HongV19} review algorithms, infrastructures and architectures for resource management on edge/fog computing. Infrastructure refers to hardware, software and middleware. The architectures are divided into data flow, control and tenancy ones and they present data flow architectures that provide techniques which take into consideration bandwidth, latency, energy efficiency, quality, security and heterogeneity in order to reduce the communication cost. Regarding the algorithms, they divide them into four categories; discovery, benchmarking, load-balancing and placement. Discovery algorithms aim to identify edge resources to use for deployment of applications. Benchmarking focuses on capturing the actual performance (CPU, memory, storage) using stress tests and tools. Load-balancing aims to distribute tasks using several approaches, such as  optimization techniques, graph-based balancing, breadth-first search or cooperative load balancing. Finally, placement algorithms assign tasks onto resources and can be divided into dynamic, condition-aware and iterative ones.

The placement techniques in \cite{HongV19} are summarized as follows. Wang et al. \cite{WangUHCZL17} propose off-line and on-line techniques for finding an optimal placement of task (service) instances on Mobile Micro Clouds in order to minimize the average cost while predicting future cost parameters with known accuracy. The off-line algorithm finds the optimal placement of services within a predefined time-window whereas the on-line approximation one makes placement decisions upon arrival of the services. Taneja et al. \cite{TanejaD17} map IoT application jobs on fog resources by ordering both based on their CPU, RAM and bandwidth requirements and capacities, respectively. In case, it is not feasible to map jobs to fog nodes, these are delegated to cloud resources.  Souza et al. \cite{SouzaMMSGRJF18} propose three algorithms, namely First-Fit, Best-Fit and Best-Fit with Queue, for placing services, which are divided into smaller atomic-services that form DAGs, on fog or cloud nodes. First-fit assigns services to the first available fog node, Best-fit first orders the services based on their requirements and then assigns them. Both these algorithms assign services to the cloud, in case no available fog node exists, whereas Best-fit with Queue puts these services in queues while waiting for fog nodes to become available by predicting if that will be more beneficial. Fernandoet al. \cite{FernandoLR19} propose Honeybee, a model for load balancing in a group of heterogeneous mobile nodes by using the work-stealing method. The independent jobs are pulled by the mobile sources instead of being pushed to them and, additionally, a node can ``steal" jobs from another node. Also, this work supports the mobility of the devices, in the sense that they can join or leave the group at any time. Skarlat et al. \cite{SkarlatNSD17} place IoT services on fog nodes taking into account their QoS requirements (e.g. execution deadlines). They organize the fog devices, called fog cells, into fog colonies, which act as data centers. Each colony has a control node used for management and communication with the other colonies and the cloud. The placement problem is formulated as an Integer Linear Programming one. Ni et al. \cite{NiZJYY17} propose a dynamic resource allocation strategy for fog computing based on Priced Time Petri Nets (PTPN) in order to improve the resource utilization while satisfying the users' QoS requirements. The authors also propose algorithms to predict the time cost and price cost of the tasks as well as compute the credibility evaluation of resources. The users can select the resources from a set of provided resources based on these characteristics. All these proposals do not account for multiple compute nodes co-operating for processing the same task in complex DAGs.

In a  second survey, Beaumont et al. \cite{BeaumontCELMMST20} present and compare algorithms that focus on scheduling tasks on CPU and GPU resources. The scheduling algorithms decide which task to execute first, when to start the execution and where to allocate it. They provide performance guarantees in the form of lower makespan bounds. They categorize the algorithms as on-line and off-line and the tasks as independent or with precedence constraints. In the on-line scenario they prove  lower bound conditions, while, in the off-line
algorithms, a linear programming problem is solved to keep the workload on CPU and GPU processors under a defined threshold and also to satisfy any precedence relations between tasks. Nevertheless, these solutions cannot be transferred to massively parallel (streaming) settings, while also heterogeneity of inter-node communication speeds are not considered. On the other hand, online scheduling techniques can inspire adaptive resource placement policies in the context of our project.

\begin{figure}[tb!]
\centering
        \includegraphics[totalheight=8cm]{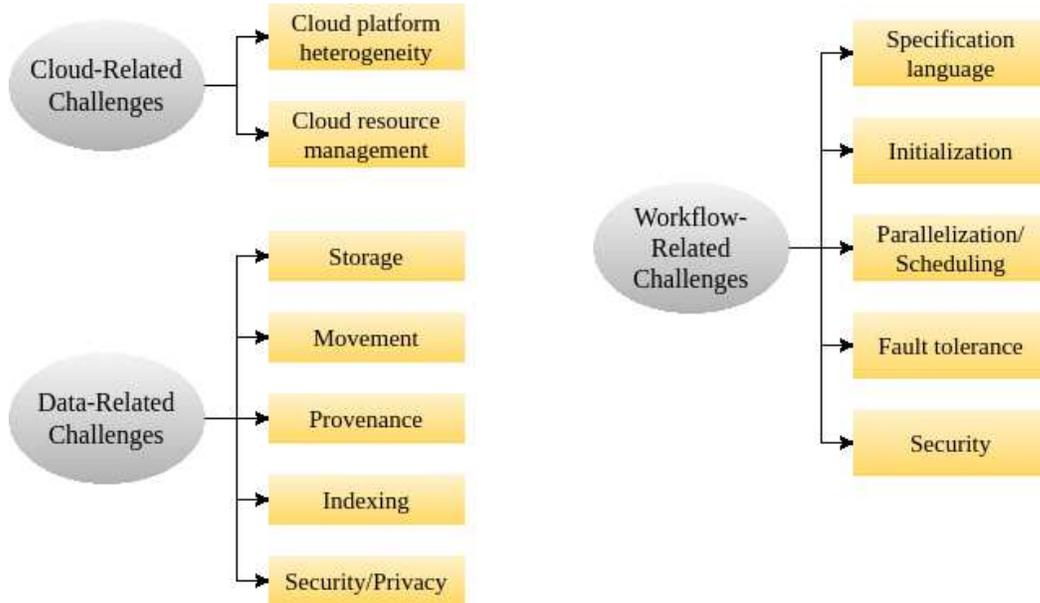}
    \caption{Workflow orchestration challenges taxonomy proposed in \cite{BarikaGZWMR19} .}
    \label{fig:BarikaTaxonomy}
\end{figure}

The third survey considered is authored by Barika et al. \cite{BarikaGZWMR19}, who review the challenges in orchestrating big data analysis workflows. First, they present the different programming models that exist, namely MapReduce, Stream, NoSQL, SQL, Message Ingestion and Hybrid. They also provide a taxonomy for the challenges, which can be divided into three categories: cloud-based, data-based and workflow-based ones as depicted in Figure \ref{fig:BarikaTaxonomy}. Next, they focus on some of the current approaches and techniques for the workflow-based challenges. In a nutshell, according to their discussion, the workflow specification language can be generic or custom with the latter limiting the ease to use multiple execution environments. The initialization of workflow execution aims to partition the workflow into fragments. This can be done by taking into consideration constraints like security, privacy, storage and compute capacities, data transfer constraints, or a combination of them or taking into consideration only the task and data dependencies within the workflow. Parallelization can be coarse-grained focusing on workflow level (when dealing with multiple workflows) or fine-grained focusing on activity/task level. Scheduling can be push-based, which aims to schedule tasks to available workers or pull-based, where the workers request for the tasks they will process. The former can be further categorized into static, dynamic or hybrid. Workflow fault-tolerance can be achieved either through reactive or proactive techniques. Finally the security challenge can be tackled by using multi-cloud architectures or replication-level techniques. No cost models are explicitly discussed, but the significance to cover issues such as resource, storage and data transmission characteristics is mentioned.

Adhikari et al. \cite{AdhikariAS19} survey scheduling strategies for workflows in cloud environments. They offer equations and definitions for multiple optimization objectives. The execution time of a task is equal to $ET_{ki}=\frac{SZ_k}{CP_i}$ where $SZ_k$ is the size of the task and $CP_i$ the capacity of the cloud VM instance which is the product of the number of cores and the size of each core (in MIPS or MFLOPS). The completion time of a task is either equal to its execution time or equal to the sum of its execution and waiting times. The waiting time for a task is 0 or equal to the maximum completion time of its predecessor tasks. The execution cost of a task refers to the cost of running on a VM instance and is equal to $\frac{ET_{ki}}{\tau}*CO_i$ where $\tau$ is the time period the task used the VM and $CO_i$ represents the cost of the VM. The energy consumption of a task refers to the amount of energy required to run on a VM instance and is equal to $E_{ik}=DP_{ik}+SP_{ik}$, that is the sum of the dynamic and static power consumption. The makespan (total execution time) of a workflow is the maximum completion time of its tasks, the total execution cost is the sum of the execution costs of all the tasks and similarly the total energy consumption is the sum of the energy consumption of all the tasks. Some of the most common scheduling objectives are total execution time (makespan), total execution cost, schedule length ratio (SLR-the ratio of the completion time of a task to the total execution time of the workflow), availability, reliability, energy consumption, resource utilization, communication overhead and security. The authors also discuss some of the emerging trends in workflow scheduling which include containers (e.g. Docker, Kubernetes), server-less computing (DynamoDB, Amazon SNS, Google cloud pub/sub) and fog computing. The latter environment also raises the need to consider the latency objective between two compute nodes $i$ and $j$ ,which is equal to $LT_{i,j}=PD_{i,j}+SD_{i,j}$. $PD_{i,j}=\frac{DS_{i,j}}{BW_{i,j}}$ is the propagation delay and depends on the distance between the nodes $DS_{i,j}$ and the bandwidth of the network $BW_{i,j}$. $SD_{i,j}=\frac{SZ_{i,j}}{TR_{i,j}}$ is the serialization delay and depends on the size of the workflow in bits and the transmission rate of the network. However, the content of this survey is not suitable for our setting, because it does not account for placement for complex streaming applications, where all DAG nodes are active simultaneously and partitioned parallelism is employed. 

Finally, Salaht et al. \cite{SalahtDL20} survey techniques for the service placement problem in Fog Computing environments. They characterize the works as being centralized or decentralized, off-line or on-line, static or dynamic and as to whether they support mobility of the devices or not. Furthermore, they focus on the optimization problems (multi- vs. mono-objective), the metrics considered (latency, resource utilization, cost, energy consumption, congestion ratio, etc.), the formulation of the problem (integer linear programming, integer non-linear programming, mixed integer linear programming, mixed integer quadratic programming) and finally the resolution strategies (exact solution, heuristics, meta-heuristics, approximations). In terms of modeling, the constraints to be covered include resource, network and applications (e.g., locality of a task placement) ones. The main limitation of the techniques examined that account for complex DAGs to be applicable to our setting is that they do not consider massive parallelism.

\newpage
\section{The Proposed Cost Model Proposal}
\label{sec:our-cost-model}

The cost model we propose is tailored to streaming data scenarios focusing on geo-distributed, heterogeneous resources. The heterogeneity apart from network speeds also refers to hardware characteristics, like CPU and RAM. Also we consider the massive parallelism of the operators. Since it refers to a streaming case, in terms of performance, the focus is on latency. The cost model extends the one presented in \cite{GuZGXH16} which however, apart from smaller differences, does not cover massive parallelism on top of arbitrary resources.

\begin{table}[b!]
  \centering
    \begin{tabular}{l|p{0.77\linewidth}} 
      \textbf{Symbol} & \textbf{Meaning} \\
      \hline
      $G_{op}$ & Graph representing the overall analytics job \\
      $V_{op}$ &Vertices of $G_{op}$ (operators or tasks)  \\
      $E_{op}$ &Edges of $G_{op}$ (data shuffling)  \\
	  $ED$ &Edge devices  \\
	  $s_{i}$ &Selectivity of $i \in V_{op}$ \\
	  $available_{i,u}$ &availability of $u \in ED$ for $i \in V_{op}$  \\
	  $comCost_{u,v}$ &Communication cost between $u,v \in ED$\\
$ED_{i}\subset ED $ &Subset of edge devices $i \in V_{op}$ can be assigned to \\
$x_{i,u}$ &Fraction of $i \in  V_{op}$ assigned on $u \in ED$ \\
	$\alpha$ & Network congestion factor \\
	$enabledLinks_{i,j}$ & Number of devices that participate in $(i, j) \in E_{op}$ \\
    \end{tabular}
    \caption{Notations used in the proposed cost model.}
\label{tab:notations}
\end{table}

Table \ref{tab:notations} presents the main notation used. Specifically, we represent a streaming analytics job as a Directed Acyclic Graph (DAG) $G_{op}=(V_{op},E_{op})$ where each node in $V_{op}$ represents an operator and the edges in $E_{op}$ represent the data flow between operators. An operator represents a set of analysis job steps  at a fine level of granularity that can run on the same device (typically in a pipelined manner). Each operator can be physically partitioned into multiple instances, where each instance is responsible for a disjoint data partition.  Edges represent data re-distribution among the partitioned operator instances. For example, according to the above description, in Apache Spark, operators correspond to Spark job stages while edges are placed when data shuffling takes place. The data sources are the operators in the $G_{op}$ without incoming edges, and we assume that they produce data in batches periodically.

Each operator takes as input multiple tuples (i.e., records) in batches that need to be analyzed and also has a certain selectivity $s_{i}$ according to its functionality. For example, a transformation operator with selectivity $s_{i}=1$ outputs a tuple for each tuple it takes as input, while a filtering operator with $s_{i}=0.5$ outputs a tuple for every two input tuples on average. If operator $i$ is a sink node in the graph, then, there is no selectivity, whereas,  if it is a source node, its selectivity is equal to 1. Each pair of edge devices has a different communication cost, $comCost_{u,v}$ ($u,v \in ED$), which relates to the connected devices network speed; the latter is also dependent on the physical distance and link capacity between the devices.

Each operator is assigned to multiple edge devices that run in parallel. That means that each device $u$ is assigned a fraction $x_{i,u}$ of tuples of an operator $i$ to analyze. However, due to privacy, security and other reasons, there exists a subset of edge devices where an operator can be assigned to (denoted by the flag $available_{i,u}$ for operator $i$ regarding device $u$). We define as $ED{i} \subset ED$ the subset of edge devices operator $i$ can be assigned to. 

As already stated, the QoS metric we try to minimize is the average latency. This latency is equal to the latency of the critical path (i.e., the slowest path) with regards to a single input data batch and consists of the average communication latency between the operators in the critical path. In order for the critical path's latency to be representative of the whole graph's latency, we assume there is no waiting time between the tasks of the operators. This is achieved through setting the operator requirements at such levels, so that they sustain the input data production rate. A natural consequence of this is that each operator can start its execution as soon as it receives its input batch of tuples from its parent nodes in the graph. As our focus is on geo-distributed realms, we make the realistic assumption that the execution latency of each operator is negligible and the communication cost dominates. The total latency of a tuple is the time to flow from its source node downstream to a sink node. 

The communication latency between two nodes is expressed by $max \{x_{i,u}*s_{i} *\sum_{v \in ED_j}\\ comCost_{u,v}*x_{j,v})\}, ~u \in ED_i$ across all instances of operator $i$. This is equal to the slowest data transfer that comes from a single device and refers to the cost of a batch of input data. However, this modeling is over-simplistic because it does not take into account the overhead of an operator instance maintaining multiple remote connections. In order to take into account such an overhead, we introduce a parameter, notated as $\alpha$ multiplied by the number of enabled links. A link between device $u$ and device $v$ for two operators $i,j$ is enabled when $x_{i,u} \neq 0$, $x_{j,v} \neq 0$ and $u \neq v$. Thus, $enabledLinks_{i,j}$ denotes the number of devices that exchange data between two operators over the network. The total latency of the topology is equal to the latency of the critical path, that is slowest path of the graph that leads from a source node to a sink one (not including). We denote as $path$ any path from a source to an operator just upstream a sink operator ($j$ is the operator just after $i$).

\begin{equation*}
Latency=max_{path\in G_{op}}\{\sum_{i\in V_{op}\in path} 
max \{x_{i,u}*s_{i}*\\\sum_{v \in ED_j} ( comCost_{u,v}*x_{j,v})+\alpha*enabledLinks_{i,j}\}\} 
\end{equation*}

\subsection{Accounting for Additional Objectives}

The above cost model targets a single objective. It is trivial, through simple sum functions to consider the movement of data over the network, as in \cite{MichailidouG19}, or the total time resources are occupied for processing a specific time period.

An important novelty, however, in modern dataflows if their focus on additional objectives, such as quality optimization. The quality of the data is an important aspect in IoT scenarios. Low quality can lead to less useful results. The quality of data can be categorized into multiple dimensions. Some examples of these are \emph{completeness, timeliness} and \emph{accuracy}. Some of the most common factors that lead to  decreased data quality include the heterogeneity of data sources, missing and dirty data due to network malfunctions or security constraints \cite{KarkouchMM+16}. Consider the scenario where a user wants to analyze the output of a sensor every 5 minutes. If, for some reason, the sensor malfunctions or restarts, the output will be inaccurate or misleading and the user should ignore it. Thus there is a need to ``rate" the quality of the data and/or to detect outliers and missing values that could lead to misleading results. However, this data quality rating can be very time consuming especially when dealing with streaming big data that need to be processed under tight latency constraints. 

$DQ_{fraction}$ denotes the percentage of the input data on which the quality check is applied. Then, we can combine the quality and latency metrics in a single formula as follows:

\begin{equation}\label{eq:8}
F=\frac{Latency}{1+\beta*(DQ_{fraction})},~\beta \geq 0
\end{equation}

The  latency represents the average time a tuple needs to come out of a sink node, beginning from a source one. The latency is increased with more data quality checks but the relationship is not proportional. This is due to the fact that the more the quality checks, the less an edge device can be assigned tasks of upstream operators, thus inducing higher communication cost.
The configuration parameter $\beta$ is a weight denoting the importance of data quality (DQ). The higher its value, the more beneficial (in terms of minimizing $F$) becomes to increase $DQ_{fraction}$. Setting $\beta$ to 0 essentially removes DQ from the optimization criteria.

\emph{Example.}
Lets assume we have a simple linear DAG with 3 nodes (operators) and 3 devices. The selectivity of each node is the following: $s_{0}=1,s_{1}=1.5$ ($s_2$ does not have any impact). The communication costs between devices are presented in Table \ref{exdetails1}. For simplicity we set $\alpha$ equal to 0.

\begin{table}[tb!]
\centering
\begin{tabular}{c|ccc}
Device & 0   & 1   & 2 \\ \hline
0      & 0   & 1.5 & 2 \\
1      & 1.5 & 0   & 1 \\
2      & 2   & 1   & 0 \\ \hline
\end{tabular}
\caption{Communication cost between edge devices in GBps.}
\label{exdetails1}
\end{table}


\begin{table}[tb!]
\centering
\begin{tabular}{c|ccc}
operator/device & 0   & 1   & 2   \\ \hline
0               & 0.8 & 0.2 & 0.0 \\
1               & 0.7 & 0.0 & 0.3 \\
2               & 0.3 & 0.4 & 0.3 \\ \hline
\end{tabular}
\caption{Fraction of operator assigned to each edge device.}
\label{fracassign1}
\end{table}

We assign the operators as shown in Table \ref{fracassign1}. The transfer time for each link in the graph depends on the selectivity of that operator, the communication cost of the devices and the assigned fractions. Each device sends data to every other device based on the fraction of the current and the succeeding operator. For example, for operator 0, device 0 will take $0.8 \times 1 \times 0 \times 0.7$ sec/unit to send data to device 0 (itself), $0.8 \times 1\times 1.5\times 0$ sec/unit to device 1 and $0.8\times 1\times 2\times 0.3$ sec/unit to device 2. We sum these cost elements to derive the total cost of communication between operator 0 and operator 1 for the first device with id 0; this sum is 0.48. For device 1, we have $0.2 \times 1 \times (1.5\times 0.7 + 0 \times 0 + 1 \times 0.3)$ = 0.27 sec/unit. For device 2, the communication cost is 0. Therefore, the latency due to the link $0 \rightarrow 1$ is $max\{0.48,0.27,0\} =0.48$. Similarly, we can derive that the cost between  $1 \rightarrow 2$ is $max\{1.26,0,0.45\} =1.26$ sec/unit. Therefore, the overall latency is 1.74 sec/unit.

If we assume that the $DQ_{fraction}$ is 0.5 then with $\beta=1$, the objective function $F$ becomes 1.16. Further assume that we examine another scenario, where $DQ_{fraction}$ is 1 at the expense of moving the complete fraction $x_{2,0}$ to device 2, i.e., the last operator runs 40\% on device 1 and 60\% on device 2. Then, the latency cost of $1 \rightarrow 2$ becomes $max\{1.89,0,0.18\} =1.89$ and the complete latency is 2.37. The new value of $F$ is 1.185, i.e., despite the important increase in latency, the new plan is not better in terms of $F$. However, if we give even more weight to the data quality through increasing the $\beta$ value, e.g., setting $\beta=2$, then the initial and modified allocation yield $F$ values of 0.87 and 0.79, respectively, i.e., the trade-off in the modification has become beneficial.

\newpage
\section{Summary}
\label{sec:summary}

In this report, we first describe existing cost modeling approaches highlighting their inadequacy to cover all aspects relevant to a modern setting including heterogeneity in compute node characteristics and network connections, massive parallelism, and complex DAG-shaped streaming application execution plans. In addition, we briefly discuss tractability issues related to the optimization problems associated with cost model proposals. To cover the aforementioned gap, in Section 3, we propose a new cost model, which also accounts for data quality. 

\emph{Acknowledgment.} The research work was supported by the Hellenic Foundation for
Research and Innovation (H.F.R.I.) under the ``First Call for H.F.R.I.
Research Projects to support Faculty members and Researchers and
the procurement of high-cost research equipment grant” (Project
Number:1052, Project Name: DataflowOpt).

\newpage
\bibliographystyle{plain}
\bibliography{cost_models_bib,ref}

\end{document}